\documentclass[useAMS]{mn2e}
\usepackage{amsmath}
\usepackage{textcomp}
\usepackage{amssymb}
\usepackage[pdftex]{graphicx}
\usepackage{amssymb}

\usepackage[margin=.8in, paperwidth=8.5in, paperheight=11in]{geometry}
\usepackage{epsfig}
\usepackage{graphicx}
\usepackage{caption}
\usepackage{esint}

\begin{document}

\title{Non-Axisymmetric Line Driven Disc Winds I - Disc Perturbations}
\author[S. Dyda, D. Proga]
{\parbox{\textwidth}{Sergei~Dyda\thanks{sdyda@physics.unlv.edu}, Daniel Proga}\\
Department of Physics \& Astronomy, University of Nevada Las Vegas, Las Vegas, NV 89154 
}

\date{\today}
\pagerange{\pageref{firstpage}--\pageref{lastpage}}
\pubyear{2017}

\label{firstpage}

\maketitle

\begin{abstract}
We study mass outflows driven from accretion discs by radiation pressure due to spectral lines. To investigate non-axisymmetric effects, we use the \textsc{Athena++} code and develop a new module to account for radiation pressure driving. In 2D, our new simulations are consistent with previous 2D axisymmetric solutions by Proga et al. who used the \textsc{Zeus} 2D code. Specifically, we find that the disc winds are time dependent, characterized by a dense stream confined to $\sim 45^{\circ}$ relative to the disc midplane and bounded on the polar side by a less dense, fast stream. In 3D we introduce a vertical, $\phi$-dependent, subsonic velocity perturbation in the disc midplane. The perturbation does not change the overall character of the solution but global outflow properties such as the mass, momentum and kinetic energy fluxes are altered by up to 100\%. Non-axisymmetric density structures develop and persist mainly at the base of the wind. They are relatively small, and their densities can be a few times higher that the azimuthal average. The structure of the non-axisymmetric and axisymmetric solutions differ also in other ways. Perhaps most importantly from the observational point of view are the differences in the so called clumping factors, that serve as a proxy for emissivity due to two body processes. In particular, the spatially averaged clumping factor over the entire fast stream, while it is of a comparable value in both solutions, it varies about 10 times faster in the non-axisymmetric case.
 
\end{abstract}

\section{Introduction}

Many accretion disc systems exhibit blue shifted absorption lines, which are interpreted as evidence of outflowing gas. These outflows are likely driven from the disc, raising the question of what physical mechanism is responsible for the driving. One possibility is radiation pressure from spectral lines, due to continuum radiation from the disc scattering off resonance lines. By including line opacities, the amount of momentum transfer is enhanced by several orders of magnitude above pure electron scattering, allowing for outflows to be launched in systems with sub-Eddington luminosity (Lucy \& Solomon 1970, hereafter LS70 and Castor, Abbott \& Klein 1975, hereafter CAK). Originally, line driving, operating in the Sobolev approximation, was proposed as a driving mechanism for stellar winds in OB stars (LS70) and was successful in predicting their mass loss rates and outflow velocities (Pauldrach, Puls \& Kudritzki 1986, Friend \& Abbott 1986). Line driving has since been applied to a variety of systems, in particular those with accretion discs.   Examples of such systems include high-state cataclysmic variables (CVs), young stellar objects (YSOs) and broad absorption line quasi-stellar objects (BAL QSOs).   

Initially theoretical studies of line driven disc winds were limited in comparison to spherically symmetric stellar winds. Some 2D theoretical models introduced simplifying assumptions to render the axisymmetric disc wind problem analytically tracktable, such as simple scaling for the strength of the radiation field (Vitello \& Shlosman 1988) or a decoupling of the radial and polar angle equations of motion (Murray et. al 1995). 

These early attempts were followed by 2D simulations carried out by Pereyra, Kallman \& Blondin (1997) and Pereyra (1997) in the context of CVs. These simulations found steady state solutions but were too coarse to resolve the subsonic part of the flow near the accretion disc.  Proga, Stone and Drew (1998, hereafter PSD98) used non-uniform meshes to resolve the subsonic parts of the flow and found that line driven disc winds in CV and YSO systems launch complex, time dependent, outflows when disc radiation is comparable to or stronger than the stellar component. Proga, Stone and Drew (1999, hereafter PSD99) used a more efficient treatment of the radiation force which allowed for the optical depth to be calculated using the Sobolev prescription without approximating the velocity gradient tensor.

An important question is what structure can affect line driven winds? Substructure on small scales or ``clumpiness" can affect the flow by altering the ionization structure of the wind, via shielding outer parts of the flow  (see for example Fullerton 2011). It can further provide new \emph{time dependent}, observable features, since diagnositcs such as the IR continuum and subordinate lines like H$\alpha$ are sensitive to $\rho^2$ (Puls et al. 1993). Clumping in disc winds has been used, albeit on different length scales, to explain in part line width ratios seen in BAL QSOs (Matthews et al. 2016)   

One possibility for generating clumpiness is instabilities in the flow, such as the line deshadowing instability (LDI) inherent in line-driven winds (Lucy \& Solomon 1970; MacGregor, Hartmann \& Raymond 1979; Carlberg 1980; Owocki \& Rybicki 1984). Owocki, Castor \& Rybicki (1988) found in 1D spherically symmetric simulations that small amplitude waves in the subsonic part of the flow can be amplified by the line driving into non-linear waves in the supersonic part of the flow. The LDI does not operate however in the Sobolev approximation that is used in CAK-like treatment of line driving, leading us to explore alternative mechanisms for generating clumps.  

In PSD98 the base of the wind is time dependent and clumpy, and matter sometimes falls back to the disc rather than being expelled. We propose that these failed winds provide a mechanism whereby an initial non-axisymmetry in the disc induces a non-axisymmetry in the wind and some matter that fails to escape the gravitational potential falls back to the disc generating new non-axisymmetries and perpetuating the cycle. Here we study this effect by first simulating an axisymmetric, line driven disc wind to test our \textsc{Athena++} radiation pressure module. We compare this to a simulation where the initial condition is perturbed and study the resulting wind outflow properties, in particlar the degree of non-axisymmetry of the solution, as well as its stability.

The structure of this paper is as follows. In Section \ref{sec:numerical_methods}, we describe our numerical methods, including the implementation of the radiation force due to electron scattering and line driving in the hydrodynamics code \textsc{Athena++}. In Section \ref{sec:2D}, we describe the results of a 3D axisymmetric disc wind, which we use as a benchmark case. We compare this run in Section \ref{sec:3D} to a 3D simulation that used the same parameters, but had an initial $\phi$ dependent, vertical, subsonic velocity perturbation in the disc midplane at $t = 0$. In Section \ref{sec:discussion}, we discuss possible observational implications for these results. We conclude in Section \ref{sec:conclusion}, where we describe possible future directions of inquiry.

\section{Numerical Methods}
\label{sec:numerical_methods}
We performed all numerical simulations with the publicly available MHD code \textsc{Athena++} (Gardiner \& Stone 2005, 2008). The basic physical setup is a gravitating central object surrounded by a thin, luminous accretion disc that provides the radiation field to drive the gas that is optically thin to the continuum.  We describe our models basic equations in Section \ref{sec:basic_equations}. Our simulations required developing a module for calculating the radiation force due to electron scattering and lines, described in Section \ref{sec:radiation_force}. For ease in comparison with previously published results we used the parameters for PSD98 ``Run 2", which was their fiducial run for line driven disc winds in CVs. These parameters, along with the boundary conditions are described in Section \ref{sec:simulation_parameters}.    

\subsection{Basic Equations}
\label{sec:basic_equations}
The basic equations for single fluid hydrodynamics driven by a radiation field are
\begin{subequations}
\begin{equation}
\frac{\partial \rho}{\partial t} + \nabla \cdot \left( \rho \mathbf{v} \right) = 0,
\end{equation}
\begin{equation}
\frac{\partial (\rho \mathbf{v})}{\partial t} + \nabla \cdot \left(\rho \mathbf{vv} + \mathbf{P} \right) = - \rho \nabla \Phi + \rho \mathbf{F}^{\rm{rad}},
\end{equation}
\begin{equation}
\frac{\partial E}{\partial t} + \nabla \cdot \left( (E + P)\mathbf{v} \right) = -\rho \mathbf{v} \cdot \nabla \Phi + \rho \mathbf{v} \cdot \mathbf{F}^{\rm{rad}} ,
\label{eq:energy}
\end{equation}
\label{eq:hydro}%
\end{subequations}
where $\rho$, $\mathbf{v}$ are the fluid density and velocity respectively and $\mathbf{P}$ is a diagonal tensor with components P the gas pressure. For the gravitational potential, we use $\Phi = -GM/r$ and $E = 1/2 \rho |\mathbf{v}|^2 + \mathcal{E}$ is the total energy where $\mathcal{E} =  P/(\gamma -1)$ is the internal energy. The total radiation force per unit mass is $\mathbf{F}^{\rm{rad}}$. The isothermal sound speed is $a^2 = P/\rho$ and the adiabatic sound speed $c_s^2 = \gamma a^2$. We take a nearly isothermal equation of state $P = k \rho^{\gamma}$ where $\gamma = 1.01$.   We can compute the temperature from the internal energy via $T = (\gamma -1)\mathcal{E}\mu m_{\rm{p}}/\rho k_{\rm{b}}$ where $\mu = 0.6$ is the mean molecular weight and other symbols have their usual meaning.

\subsection{Radiation Force}
\label{sec:radiation_force}
We assume an axisymmetric, time independent radiation field is provided by a geometrically thin accretion disc along the midplane. The frequency integrated intensity is
\begin{equation}
I(r_d) = \frac{3}{\pi} \frac{GM}{r_*^2}\frac{c}{\sigma_e} \Gamma_d \left( \frac{r_*}{r_d} \right)^3 \left[ 1 - \left(\frac{r_*}{r_d} \right)^{1/2}\right], 
\label{eq:intensity}
\end{equation}  
where $r_d$ is the radial position on the disc, $r_*$ is the inner radius of the disc, $\sigma_e$ is the Thompson cross section per unit mass, $c$ the speed of light and 
\begin{equation}
\Gamma_d = \frac{\dot{M}_{\rm{acc}} \sigma_e}{8 \pi c r_*},
\end{equation}
is the disc Eddington number, where $\dot{M}_{\rm{acc}}$ is the accretion rate in the disc (e.g., Pringle 1981). In addition, we assume the central object of radius $r_*$ is optically thick and is not a source of driving radiation. This affects the radiation field at small radii, as the central object effectively shadows the wind from the radiation from the inner disc. 

We assume all gas is optically thin to continuum radiation and every point in the wind experiences a radiation force
\begin{equation}
\mathbf{F}^{\rm{rad}} = \mathbf{F}^{\rm{rad}}_e + \mathbf{F}^{\rm{rad}}_{L},
\end{equation}  
which is a sum of the contributions due to electron scattering $\mathbf{F}^{\rm{rad}}_e$ and line driving $\mathbf{F}^{\rm{rad}}_{L}$. In this continuum optically thin approximation, the radiation force due to electron scattering is
\begin{equation}
\mathbf{F}^{\rm{rad}}_e = \varoiint_D \left( \mathbf{n} \frac{\sigma_e I d\Omega}{c} \right),
\label{eq:f_e}
\end{equation} 
where $\mathbf{n}$ is the normal vector from the disc to the point in the wind, $d\Omega$ is the solid angle spanned by the disc and the integration is carried out over the entire disc. We note that the above integrand is time independent and for each gridpoint in the wind the integration can therefore be carried out before the start of the hydrodynamics simulation.

We treat the radiation due to lines using a modification of the CAK formulation where the radiation force due to lines is 
\begin{equation}
\mathbf{F}^{\rm{rad}}_L = \varoiint_D M(t) \left( \mathbf{n} \frac{\sigma_e I d\Omega}{c} \right),
\label{eq:f_line}
\end{equation}
and $M(t)$ is the so-called force multiplier. The force multiplier parametrizes how many lines are effectively available to increase the scattering coefficient. We use the Owocki, Castor \& Rybicki (1988, hereafter OCR) parametrization of the line strength, where working in the Sobolev approximation, it is a function of the optical depth parameter
\begin{equation}
t = \frac{\sigma_e \rho v_{\rm{th}}}{|dv_{l}/dl|},
\label{eq:optical_depth_parameter}
\end{equation} 
where $v_{\rm{th}}$ is the thermal velocity of the gas and the velocity gradient along the line of sight
\begin{equation}
\frac{dv_l}{dl} \equiv Q = \epsilon_{ij} n_i n_j,
\label{eq:full_Q}
\end{equation}
can be expressed as a sum over elements of the shear tensor $\epsilon_{ij}$ where $n_i$ are the components of the normal vector. The OCR formulation for the force multiplier is
\begin{equation}
M(t) = k t^{-\alpha} \left[ \frac{(1 + \tau_{\rm{max}})^{1-\alpha} - 1}{\tau_{\rm{max}}^{1 - \alpha}}\right],
\end{equation} 
where $k$ and $\alpha$ are constants, $\tau_{\rm{max}} = t \eta_{\rm{max}}$, and $\eta_{\rm{max}}$ is related to the maximum force multiplier via $M_{\rm{max}} = k (1 - \alpha) \eta_{\rm{max}}^{\alpha}$. This force multiplier parametrization follows CAK for large optical depths but saturates to $M_{\rm{max}}$ as optical depth becomes very small. 

Stevens and Kallamn (1990) used detailed photoionization calculations to determine the radiation force driving the stellar wind in a massive X-ray binary (MXRB).  They found the ionization state depends on the SED, in particular the hardness of the X-rays. Line driven wind simulations of QSOs (Proga, Stone \& Kallman 2000; Nomura et al 2016) have accounted for this in a self-consistent way by including the heating/cooling primarily due to X-rays. Our simulations apply to CVs, where the X-ray flux is softer than in QSOs. We therefore approximate the flow as isothermal and leave careful modeling of the wind ionization structure for future work.

The radiation force due to lines (see eq. \ref{eq:f_line}) has an integrand that is time dependent, because it depends on the velocity gradient and the gas density through the optical depth parameter (see eq. \ref{eq:optical_depth_parameter}). This is computationally expensive because the integration over the disc must be carried out at every time step. PSD98 assumed that the dominant contribution to the velocity gradient is along the vertical direction, 
\begin{equation}
\begin{split}
Q \approx \frac{dv_z}{dz} n_z^2 =  \left[ \frac{\partial v_r}{\partial r} \cos^2 \theta + \frac{1}{r} \left( \frac{\partial v_{\theta}}{\partial {\theta}} + v_r \right) \sin^2 \theta \right.  \\ \left. + \left( \frac{v_{\theta}}{r} - \frac{\partial v_{\theta}}{\partial r} - \frac{1}{r}\frac{\partial v_r}{\partial \theta} \right) \sin \theta \cos \theta \right] n_z^2.
\end{split}
\end{equation}  
The radiation force due to lines is then
\begin{equation}
\mathbf{F}^{\rm{rad}}_L = M(t) \varoiint_D  \left( \mathbf{n} \ n_z^{2\alpha} \frac{\sigma_e I d\Omega}{c} \right),
\label{eq:f_line_dvdz}
\end{equation}
where we have assumed that $\tau_{\rm{max}} = \sigma \rho v_{\rm{th}}/|dv/dz| \eta_{\rm{max}}$ (see appendix C in PSD98 for a derivation). Since all time dependent factors are in $M(t)$, the integration over the disc must only be performed at the start of the simulation. PSD99 found that this approximation gave good agreement with simulations using the full expression for Q (eq. \ref{eq:full_Q}) so we make use of it here. We describe our numerical integration scheme, based on PSD99, in more detail in Appendix \ref{sec:appendix}.

\subsection{Simulation Parameters}
\label{sec:simulation_parameters}
We chose parameters that correspond to PSD98 Run2. The central object has mass and radius $M = 0.6 \ M_{\odot}$ and $r_* = 8.7 \times 10^{8} \ \rm{cm}$, respectively. The sound speed $c_s = 14 \ \rm{km/s}$, which corresponds to a hydrodynamic escape parameter $\rm{HEP} = GM/r_*c_s^2 = 4.6 \times 10^4$ at the base of the wind. For this high HEP, thermal driving, which requires $\rm{HEP}$ no more than $10$ (Stone \& Proga 2009; Dyda et al. 2017), is negligible throughout the domain, .

We impose outflow boundary conditions at the inner and outer radial boundaries and axis boundary conditions along the $\theta = 0$ axis. We assume a reflection symmetry about the $\theta = \pi/2$ midplane. In the $\phi$ direction, we impose periodic boundary conditions, where our 3D simulation covers a range $0 < \phi < \pi/4$. After every full time step we reset $\rho_d$ to $10^{-8} \rm{g \ cm}^{-3}$, $v_r$ to $0$ and $v_{\phi}$ to $v_K = \sqrt{GM/r}$. We also impose that the velocity $v_{\theta}$ is unchanged due to resetting density by also resetting the momentum i.e. $(\rho v)_{\theta}^{n+1} = \rho_d (\rho v)_{\theta}^{n}/(\rho)^{n}$ where the superscript refers to the $n$ and $n+1$ timestep respectively.  Generally we followed the setup of PSD98 as closely as possible, given that they used the \textsc{Zeus} 2D code (Stone \& Norman 1992).

We choose a domain size $n_r \times n_{\theta} \times n_{\phi} = 96 \times 96 \times 64$ and $N_r \times N_{\theta} \times N_{\phi} = 3 \times 3 \times 2 = 18$ MPI meshblocks of size $32^3$. Meshblocks of size 32 is the recommended smallest size for \textsc{Athena++} to achieve good scaling with number of processors. The radial domain extends over the range $r_* < r < 10 \ r_*$ with logarithmic spacing $dr_{i+1}/dr_i = 1.05$. The polar angle range is $0 < \theta < \pi/2$ and has logarithmic spacing $d\theta_{j}/d\theta_{j+1} = 1.05$ which ensures that we have sufficient resolution near the disc midplane to resolve the density length scale $\lambda_{\rho}$ of the hydrostatic equilibrium disc that forms i.e to obtain $r \Delta \theta \ll \lambda_{\rho}$. We use linear spacing in the $\phi$ direction and with this resolution expect to be able to resolve the $n_{\phi}/10 \sim 6$ highest frequency axial modes that can be excited in this wedge. 

Initially, the cells along the disc are set to have $\rho = \rho_d$, $v_r = v_{\theta} = 0$, $v_{\phi} = v_K$ and the pressure $P = c_s^2 \rho^{\gamma}$. In the rest of the domain $\rho = 10^{-20} \rm{g} \rm{cm}^{-3}$, $P = c_s^2 \rho^{\gamma}$ and $v_r = v_{\theta} = v_{\phi} = 0$.

We take the disc Eddington number $\Gamma_d$ to be $3.76 \times 10^{-4}$. The thermal velocity of the gas is $v_{\rm{th}} = 4.2 \times 10^5 \rm{cm/s}$. The line driving parameters are chosen to be $k = 0.2$, $\alpha = 0.6$ and $M_{\rm{max}} = 4400$. The disc as a source of radiation is taken to extend outside the computational domain to $R_d = 30 \ r_*$.  

We impose a density floor $\rho_{\rm{floor}} = 10^{-22} \rm{g \ cm}^{-3}$, which adds matter to stay above this floor, while conserving momentum, if the density ever drops below it. In addition, we have a pressure floor $P_{\rm{floor}} = c_s^{2} \rho_{\rm{floor}}^{\gamma}$, so that our nearly isothermal condition is preserved.

\section{Results}
\label{sec:results}
We describe in detail two 3D simulations of line driven disc winds. The first, a 3D analogue of the ``Run 2" simulation from PSD98 is used as a validation of our implementation of line driving in \textsc{Athena++} and to ensure that axisymmetry is explicitly preserved in the $Q \approx dv/dz$ approximation. These results are shown in Section \ref{sec:2D}. We then perturb the intial conditions along the midplane with a vertical, subsonic, sinusoidal, velocity perturbation. The results of this perturbed run are shown in Section \ref{sec:3D} where we compare properties to the unperturbed run and also quantify its non-axisymmetric features.

\subsection{Axisymmetric Reference Run}
We performed a 3D simulation of a line driven disc wind to test that \textsc{Athena++} preserves the axisymmetry of the initial condition and of the line driven wind. 
As a consistency check, we also compared our results to the 2D simulations of PSD98.

\label{sec:2D} 
At the start of the simulation gas enters the simulation region from the midplane, accelerated by the pressure gradient and a radiation force that progressively ramps up over the first 100 s, corresponding to 10\% of the simulation runtime. After roughly 200 s a line driven wind forms, with a wind opening angle of $\sim 45^{\circ}$. Most of the mass flux is carried in a fast stream, within a roughly $ 20^{\circ}$ wide wedge, below the upper wind envelope. Above this wedge, there is a polar funnel region with density so low that it is negligible. Below this wedge, the velocity is low, and the flow turbulent enough that the mass flux is insignificant. The wind is time dependent, though the streamlines do not change significantly in time.

\begin{figure}
                \centering
                \includegraphics[width=0.5\textwidth]{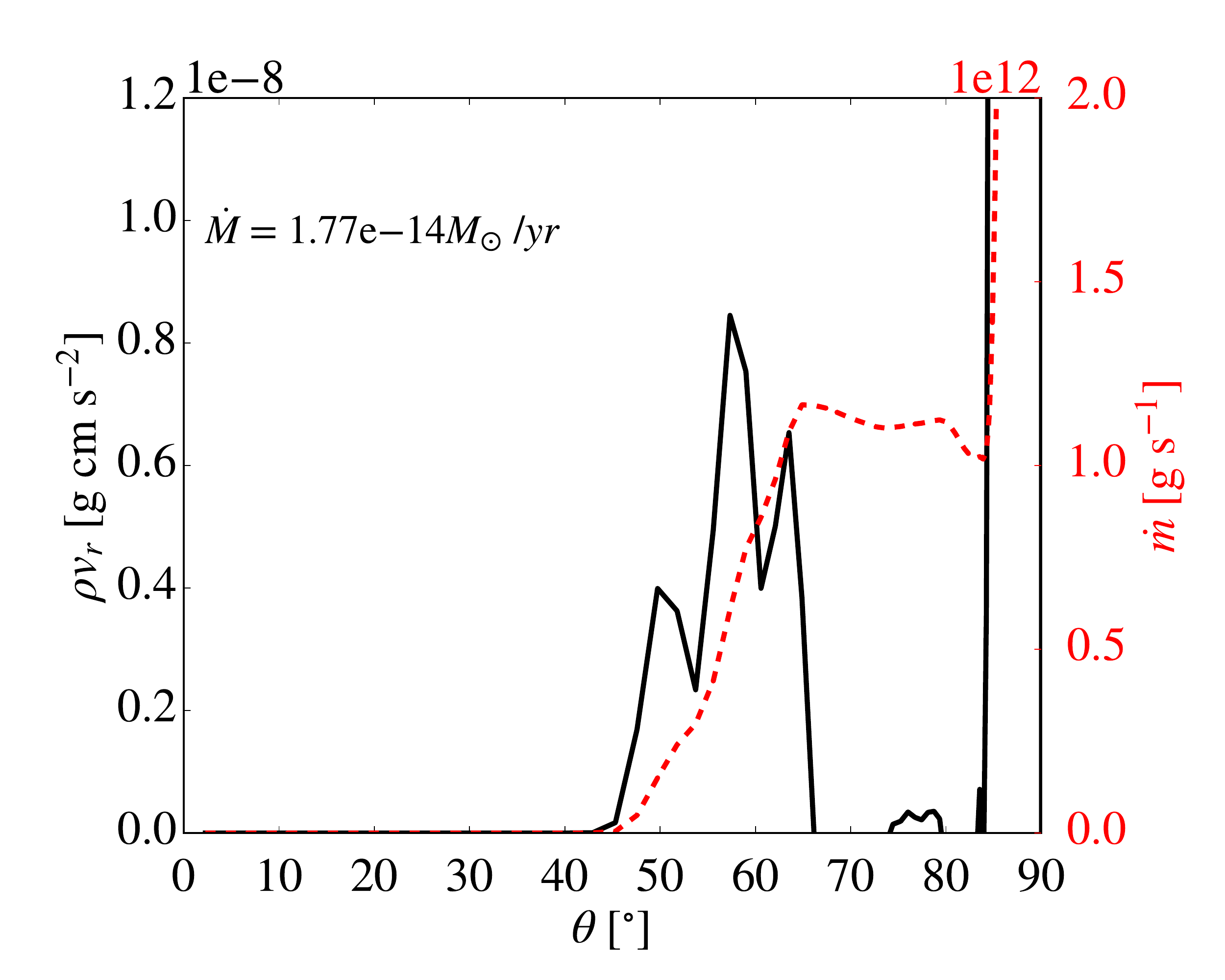}
                \includegraphics[width=0.5\textwidth]{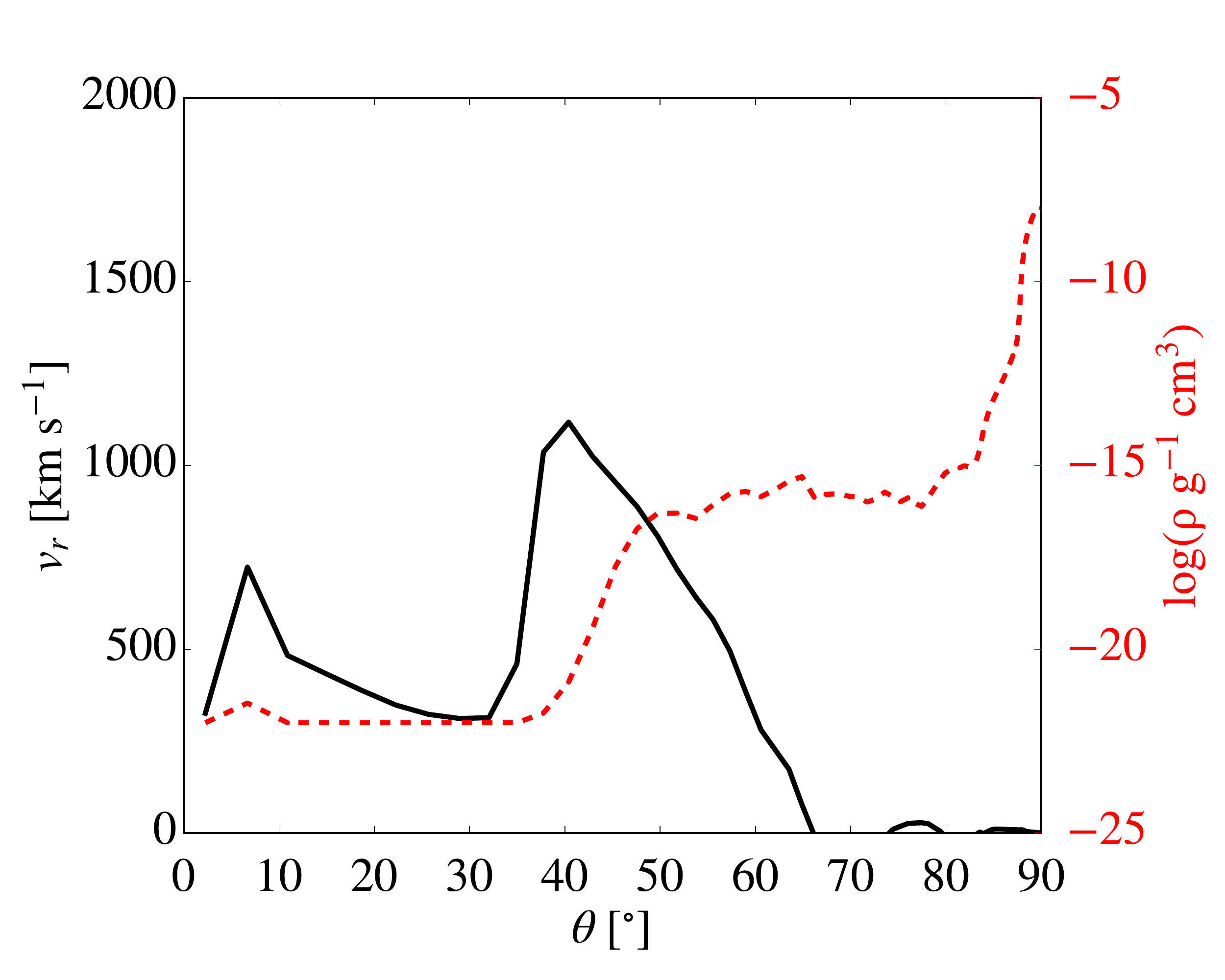}
        \caption{\textit{Top} - Momentum density $\rho v_r$ and $\theta-$integrated mass flux $\dot{m}$ along the outer boundary at $t = 1000$  s. We note that most of the mass flux is concentrated along $45^{\circ} < \theta < 65^{\circ}$. To measure wind properties, rather than disc effects we restrict our analysis of global properties to $\theta < 80^{\circ}$. \textit{Bottom} - Outflow velocity $v_r$ and density $\rho$ along the outer boundary at $t = 1000$ s. The characteristic wind speed $v_r \sim 900 \ \rm{km/s}$ and density $\rho \sim 10^{-16} \ \rm{g/cm^2}$ is consistent with 2D results from PSD98.}
\label{fig:mdot_out}
\end{figure}

The structure of the wind can be further characterized by considering its properties at the outer radial boundary. We define the $\theta-$integrated mass flux
\begin{equation}
\dot{m} = \int_0^{\theta}  \ \rho v_r \ \sin \theta d\theta.
\end{equation}
In Fig. \ref{fig:mdot_out} (top panel), we plot the momentum density $\rho v_r$ (black, solid line) and integrated mass flux (red, dashed line) as a function of $\theta$ at the outer boundary $r = 10 \ r_*$ at $t = 1000 \ \rm{s}$. We also indicate the total mass flux $\dot{M}$ from the full simulation region (see the top left corner of the top panel), assuming top/bottom symmetry. The momentum density is concentrated around $45^{\circ} < \theta < 65^{\circ}$, where the stream reaches the outer radius. This can also be seen from the integrated mass flux that levels off for larger values of $\theta$. A spike above $\theta > 85^{\circ}$ occurs near the midplane because of the much higher density disc and small subsonic $v_r$ that is positive in this snapshot. Therefore we impose a cutoff of $\theta < 80^{\circ}$ when measuring fluxes of both simulations, because we are interested in wind properties.
  
In the bottom panel of Fig. \ref{fig:mdot_out}, we plot the density and velocity structure of the wind as a function of $\theta$ at the outer radial boundary. The polar region is nearly void of matter with $\rho \sim 10^{-22} \rm{g \ cm^{-3}}$ for $\theta < 45^{\circ}$. Then there is a fast-stream region where the velocity peaks to its maximum value $\sim 1200 \ \rm{km \ s^{-1}}$ and the density of the wind increases to $\rho \sim 10^{-16} \ \rm{g \ cm^{-3}}$. The density remains nearly constant at this value for $\theta$ between $50^{\circ}$ and $80^{\circ}$.  We define the characteristic velocity of the outflow as the velocity of the wind when this reference density is reached, roughly $v_r \approx 900 \ \rm{km/s}$. We measure a wind mass loss rate to be $\dot{M} \sim 10^{-14} \ M_{\odot}/\rm{yr}$. We find our 3D solutions both confirm the results of PSD98 and maintain axisymmetry to machine precision.

\subsection{Non-Axisymmetric Perturbation}
\label{sec:3D}
The main subject of this investigation is to explore non-axisymmetric effects in line driven disc wind solutions. Non-axisymmetry can be introduced by having forces that explicitly break the axisymmetry or by perturbing the axisymmetric initial conditions. Here we want to establish the level of non-axisymmetry generated in the wind by perturbing the initial conditions. We introduce an initial vertical, subsonic, sinusoidal, velocity perturbation
\begin{equation}
\delta v_{\theta}\left|_{\theta = \pi/2} \right. = 0.1 c_s \sin (8 \phi),
\label{eq:perturbation}
\end{equation}   
in the disc midplane. We note this perturbation is the lowest mode consistent with our periodic boundary conditons in the axial direction i.e., $\delta v_{\theta}\left|_{\phi = 0} \right. = \delta v_{\theta} \left|_{\phi = \pi/4}\right.$ and $\partial_{\phi}\delta v_{\theta} \left|_{\phi = 0} \right. = \partial_{\phi}\delta v_{\theta} \left|_{\phi = \pi/4} \right.$. We compare both the global and local properties of this perturbed solution with the axisymmetric wind from the previous section.

\subsubsection{Global Properties}
\label{sec:global_properties}
\begin{figure}
                \centering
                \vspace{-1.25cm}
                \includegraphics[width=0.5\textwidth]{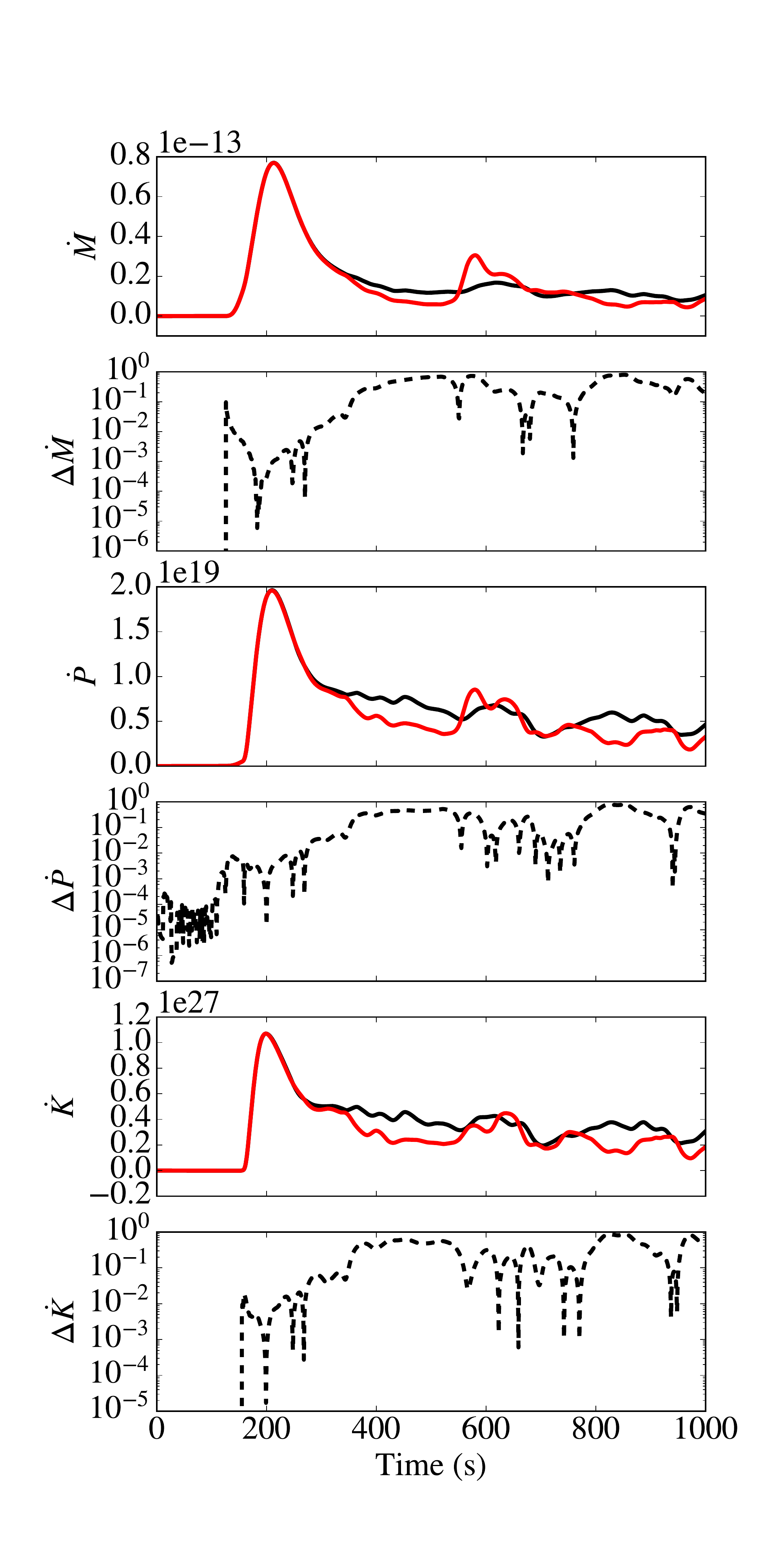}
        \caption{Mass, momentum and kinetic energy fluxes as a function of time for the unperturbed (red, solid line) and perturbed (black, solid line) solutions and the relative difference in fluxes between solutions (black, dashed line).}
\label{fig:fluxes}
\end{figure}

First, we consider the global properties of the flow. In particular, we consider the mass flux
\begin{equation}
\dot{M} = \iint  \rho v_r \ r^2 \sin \theta d\theta d\phi,
\end{equation}
momentum flux
\begin{equation}
\dot{P} = \iint  \rho v_r^2 \ r^2 \sin \theta d\theta d\phi,
\end{equation}
and kinetic energy flux
\begin{equation}
\dot{K} = \iint  \rho v_r^3 \ r^2 \sin \theta d\theta d\phi,
\end{equation}
across the outer radial boundary. We plot these quantities in Fig. \ref{fig:fluxes}, showing each of them for the unperturbed (red, solid line) and perturbed (black, solid line) solutions as well as the relative difference in fluxes between solutions (black, dashed line). At early times, the wind has not yet reached the outer boundary and fluxes are nearly zero. After an initial transient outburst at time $\sim 200 \ \rm{s}$, the solutions settle down and exhibit a relative difference of up to 100\%. We note that both solutions are highly time dependent, and there is no clear trend about which solution has a more powerful wind. However, we can conclude that the perturbation has a non-negligible effect on the gross outflow properties. 

\subsubsection{Local Deviations}
\label{sec:phi_properties}

\begin{figure*}
                \centering
                \includegraphics[width=1\textwidth]{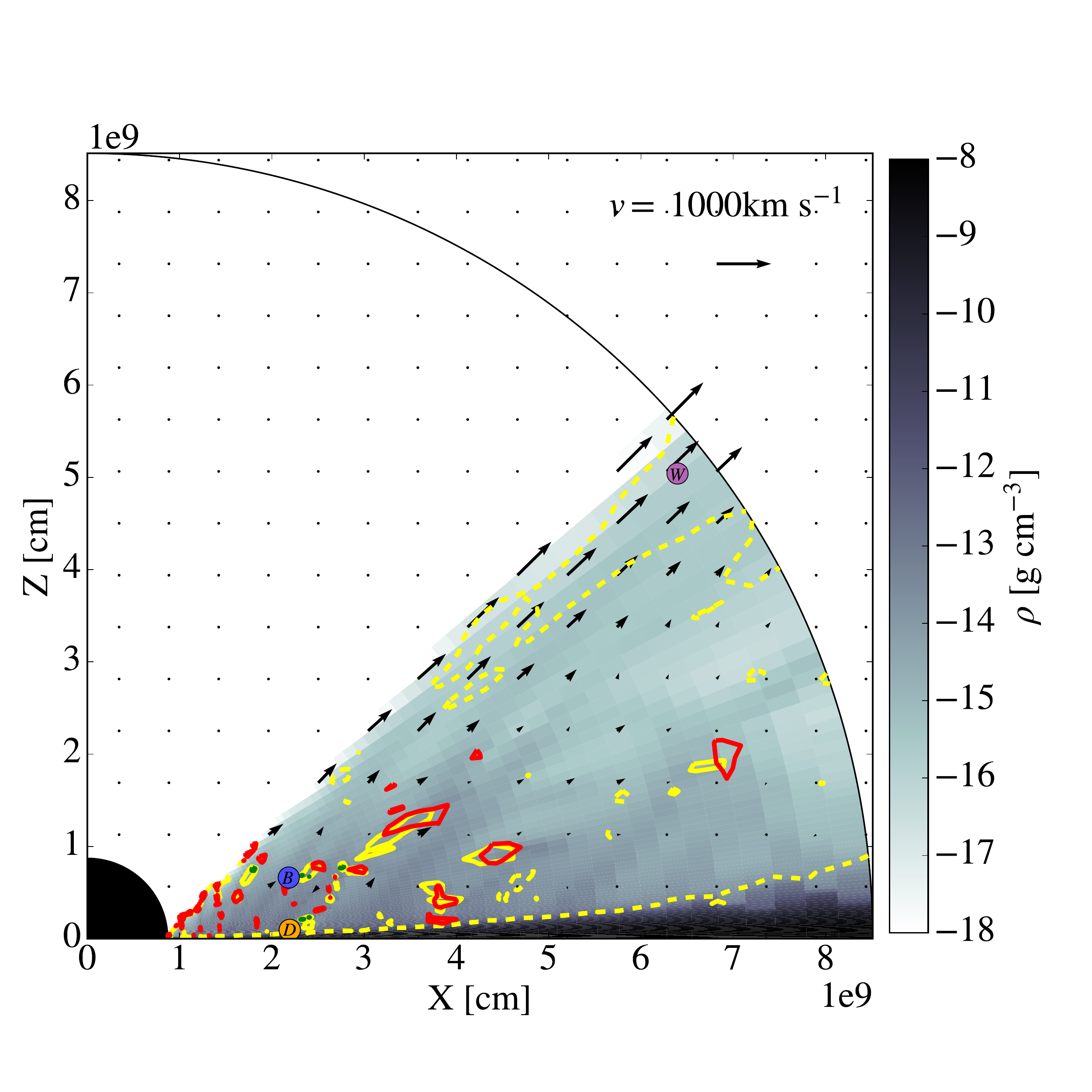}
        \caption{Azimuthally averaged density contours and velocities at t = 1000 s for the perturbed simulation. We indicate the positions where $\sigma (\rho) = 10^{-1}$ (yellow, dashed contours) and $10^0$ (yellow, solid contours). We also mark over-dense (green) and under-dense (red) contours. i.e features which differ from the azimuthal average by a factor of 3. In addition, we indicate the location of three points in each of the representative parts of the flow: the disc (orange,``D"), base of the wind (blue,``B") and the main wind itself (purple,``W"). These points lie on a single streamline (not shown) and are used later in a more detailed analysis.}
\label{fig:clumping}
\end{figure*} 

\begin{figure}
                \centering
                \includegraphics[width=0.5\textwidth]{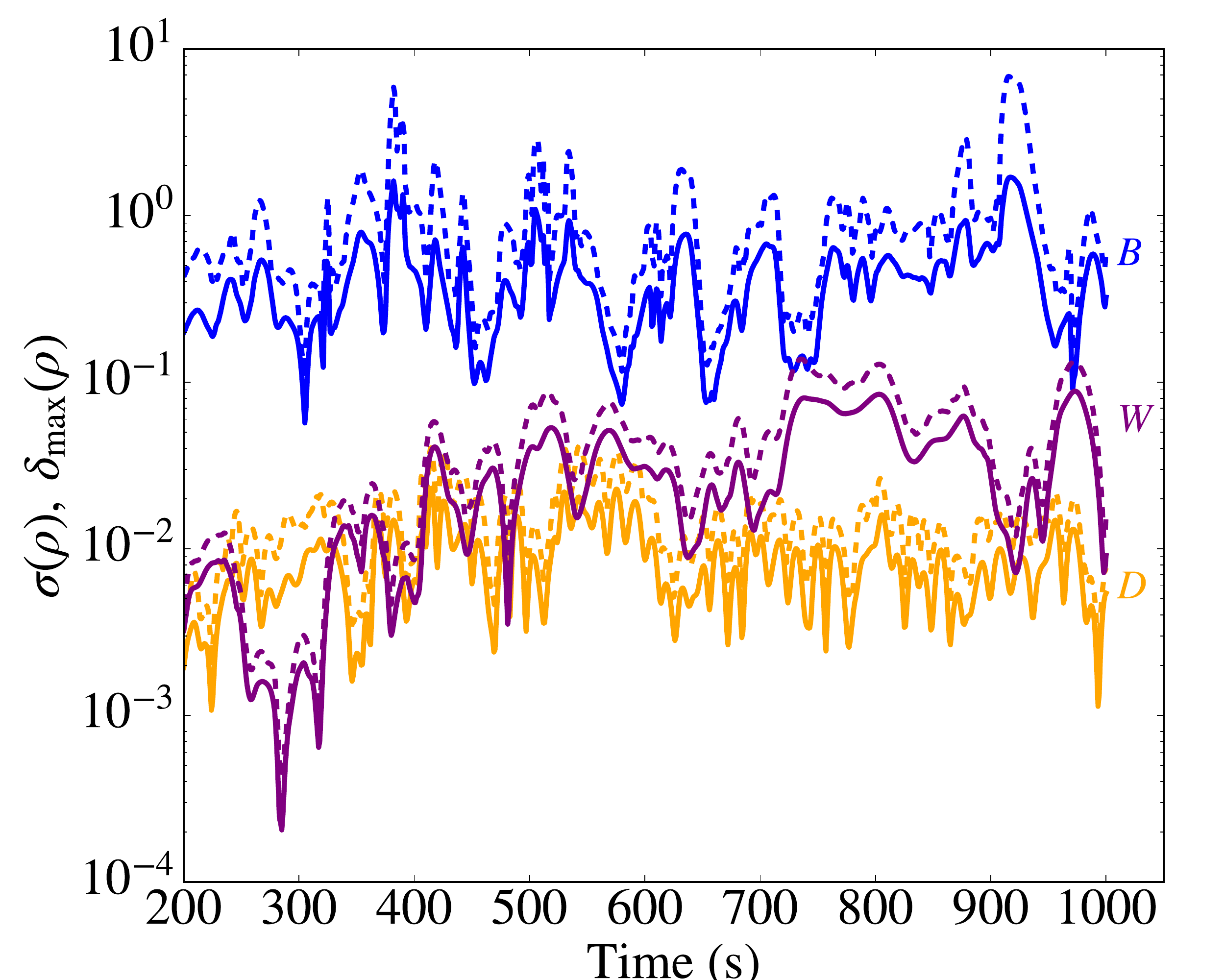}
        \caption{Relative standard deviation $\sigma(\rho)$ (solid) and relative maximum deviation $\delta_{\rm{max}}(\rho)$ (dashed) for the locations indicated by the corresponding colored points indicated in Fig. \ref{fig:clumping}. After an initial period of wind formation, the standard deviation maintains a level of $\sim 10\%$ inside the wind (purple) but saturates to a much smaller level of $10^{-3}$ inside the disc (orange). The deviation is largest at the base of the wind (blue) where it is sometimes $~ 100 \%$. The $\sigma$ and $\delta_{\rm{max}}$ distributions have the same features, indicating that the dominant contribution to the standard deviation are the locations where the deviation is maximal.}
\label{fig:deviation_2d3d}
\end{figure} 

To quantify departures from axisymmetry we define the relative standard deviation
\begin{equation}
\sigma\left( \rho \right) = \frac{1}{\bar{\rho}} \sqrt{\sum_{k = 0}^{N_{\phi}}\left( \rho_k - \bar{\rho}\right)^2},
\label{eq:stddev}
\end{equation}
where $\bar{\rho}$ is the $\phi-$averaged density distribution. This metric quantifies how density at fixed $(r,\theta)$ deviates, on average, in the $\phi$ direction. We also define the relative maximum deviation
\begin{equation}
\delta_{\rm{max}}\left( \rho \right) = \frac{1}{\bar{\rho}} \rm{max}\Big| \rho_k - \bar{\rho} \Big|.
\end{equation}
This provides a measure of the largest deviation from axisymmetry at a fixed $(r,\theta)$. Because emission scales like $\rho^2$, we consider over(under)-dense regions observationally interesting. To potentially emit $\sim 10$ times more (less) than the azimuthal average, the density must differ by a factor of about three from the azimuthal average. We therefore refer to over-dense regions where $\rho_k \geq 3 \bar{\rho}$ and under dense regions with $\rho_k \leq \bar{\rho}/3$. 

We identify three qualitatively different regions in our simulations - a nearly hydrostatic \emph{disc}, a wind \emph{base} and the main part of the \emph{wind}. To sample these regions we choose points at $(r,\theta) = (2.5 \ r_*, 89.2^{\circ})$, $(2.5 \ r_*, 84^{\circ})$ and $(9.8 \ r_*, 52^{\circ})$ inside of each of these respective regions. These points lie on a single streamline, and are used in later analyses.  

In Fig \ref{fig:clumping}, we plot the $\phi-$averaged density and velocity at $t = 1000 \ \rm{s}$. We mark the $\sigma(\rho) = 10^{-1}$ and $10^{0}$ contours (yellow, dashed and solid contours, respectively) as well as over-dense $\rho_k = 3 \bar{\rho}$ (green contours) and under-dense $\rho_k = \bar{\rho}/3$ (red contours) i.e. features which differ from the azimuthal average by a factor of 3. We indicate the location of our three representative points in the ``disc" (``D", orange), ``base" (``B", blue) of the wind and the main ``wind" (``W", purple), later used in more detailed analysis.  For clarity we have not plotted vectors and contours where $\rho < 10^{-17}$ $\rm{g/cm}^3$. 

Qualitatively, this solution resembles the unperturbed solution. The standard deviation countours indicate that non-axisymmetries are small in the disc $(\sigma(\rho) < 10\%)$, but are roughly $\sim 10\%$ (dashed yellow contours) in most of the wind. There are some small regions at the base of the wind where $\sigma(\rho) \sim 100\%$ (solid yellow contours).  

In Fig. \ref{fig:deviation_2d3d}, we plot $\sigma (\rho)$ (solid) and $\delta_{\rm{max}} (\rho)$ (dashed)  as a function of time for the ``wind", ``base" and ``disc" points in their respective color. Within the disc the effects of the perturbation are nearly constant at late time, with $\sigma (\rho) \sim \delta_{\rm{max}} (\rho)  \sim 10^{-3}$. This is also evident in Fig \ref{fig:clumping} from the location of the $10^{-1}$ contour that follows the contours of the disc. As the wind is turned on, the deviation grows to $\sigma (\rho) \sim \delta_{\rm{max}} (\rho)  \sim  10^{-1}$ at the base and remains at this level. In the fastest part of the wind, the deviation is $\sigma (\rho) \sim \delta_{\rm{max}} (\rho)  \sim  10^{-2}$, smaller than at the base by an order of magnitude. 

These results are consistent with two other numerical tests we performed with line driving turned off.  We perturbed a disc that is too cold to thermally launch outflows ($\rm{HEP} = 4.6 \times 10^{4}$), and found non-axisymmetries in the disc to be of the same order of magnitude as with line driving turned on. We also considered a disc that is hot enough to thermally launch outflows ($\rm{HEP} = 10$), and found that non-axisymmetries were restricted to the streams at the polar/wind boundary as non-axisymmetric features in the main wind were advected from the simulation region.

We note that $\sigma (\rho) \sim \delta_{\rm{max}} (\rho) $, indicating that the dominant contribution to the relative standard deviation is primarily from the point where the deviation is maximal. In particular, we note how both distributions exhibit the same time behaviour. We can see this as well from the over-dense and under-dense contours, which have considerable overlap with the $\sigma$ contours. It is also important to note that there are spatially more under-dense regions than over-dense regions. This is expected since if mass in over and under-dense regions is identical then we would expect the volume of over-dense regions to be 9 times smaller than the under-dense regions.

\subsubsection{Clumpiness}

\begin{figure}
                \centering
                \includegraphics[width=0.5\textwidth]{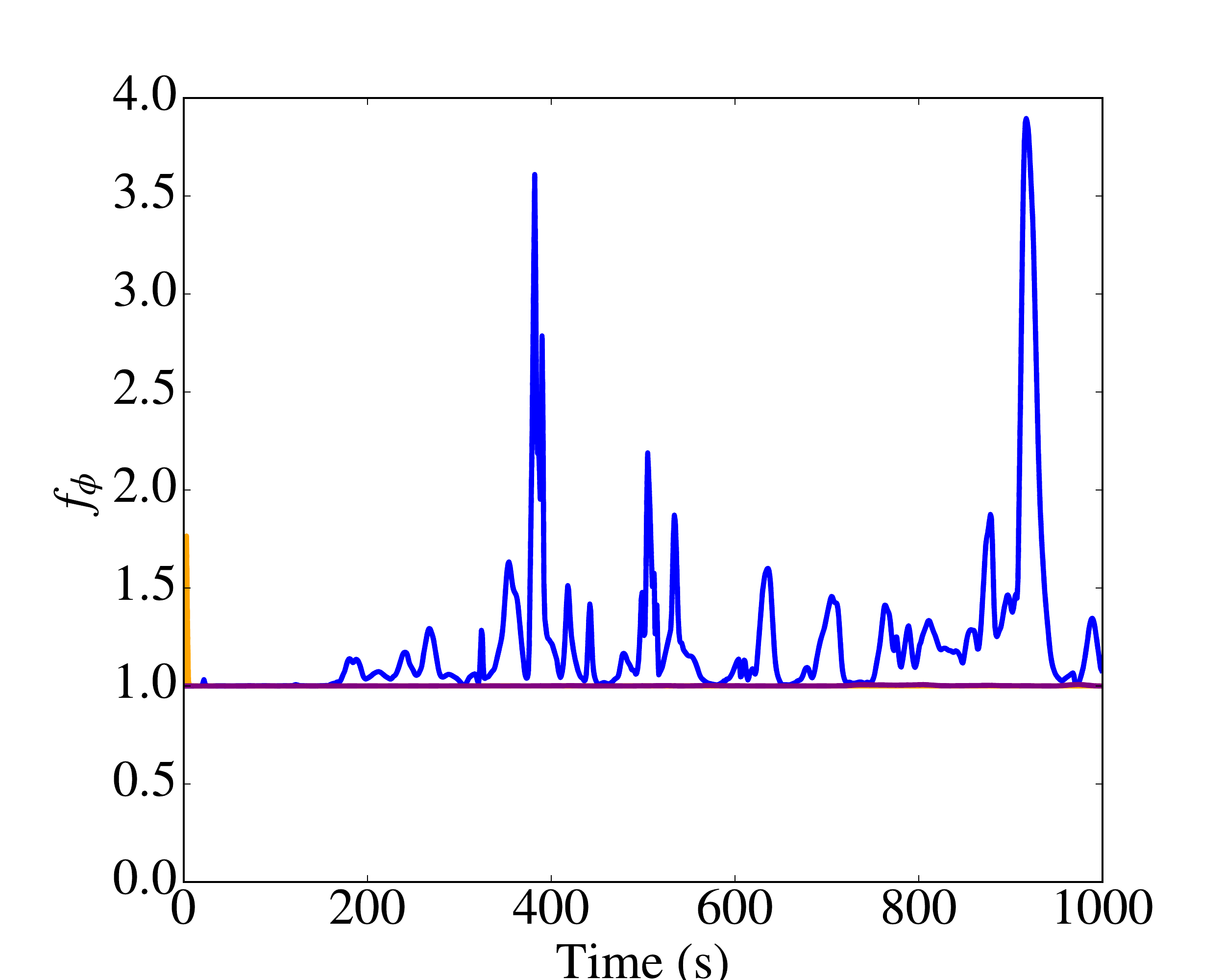}
        \caption{Azimuthal clumpiness $f_{\phi}$ as a function of time for the disc (orange), base (blue) and wind (purple) points. Azimuthal clumpiness deviates significantly from axisymmetric winds, where $f_{\phi} \equiv 1$, in the wind base.}
\label{fig:azimuthal_clump}
\end{figure} 

\begin{figure}
                \centering
                \includegraphics[width=0.45\textwidth]{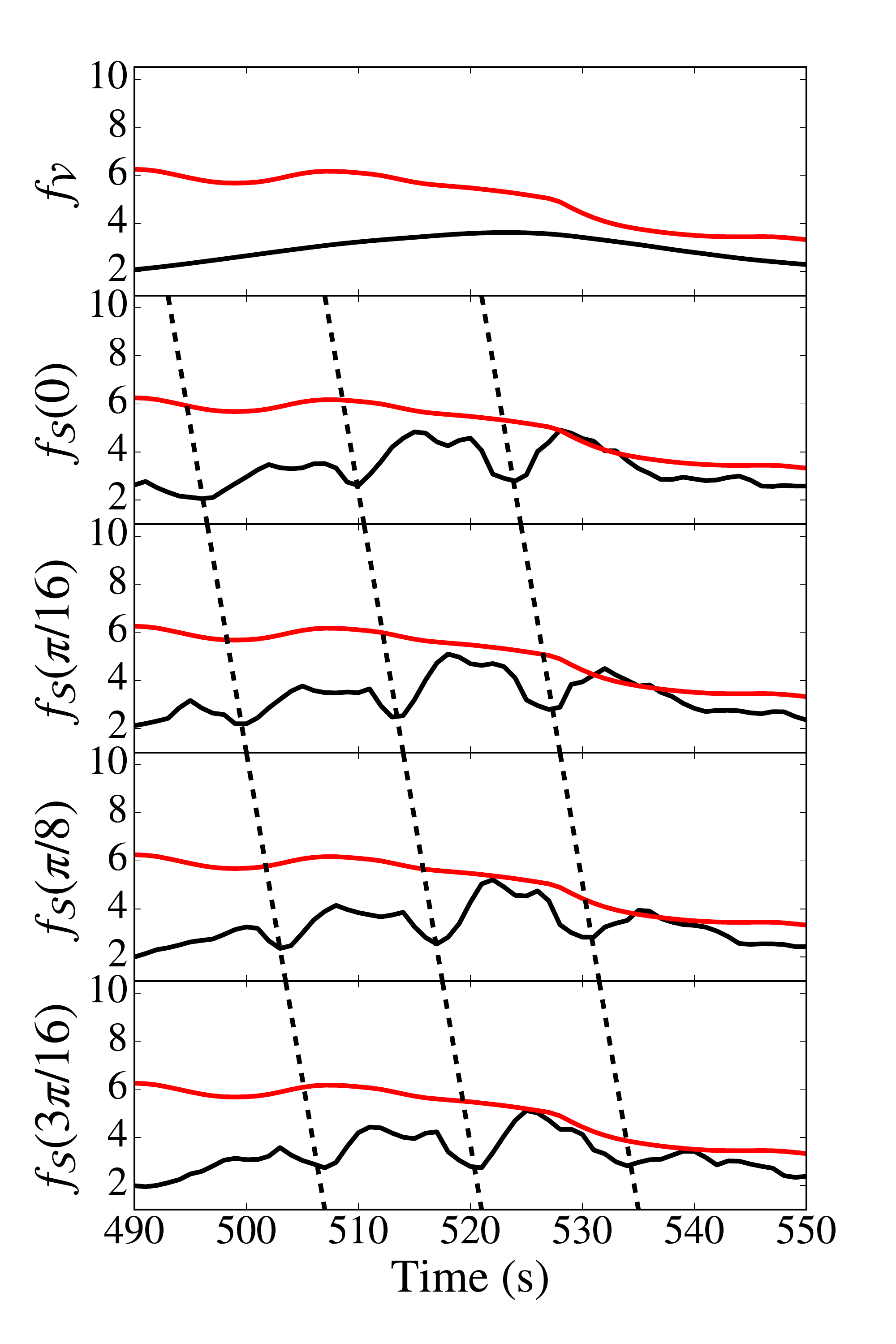}
        \caption{Volume clumping $f_{\mathcal{V}}$ (top panel) for unperturbed wind (red) and for the perturbed wind (black) and surface clumping $f_{\mathcal{S}}$ for fixed $\phi = 0$, $\pi/16$, $\pi/8$ and $3\pi/16$ (lower panels). The dashed black line indicates the motion of a feature orbiting at constant angular velocity $\Omega_{cl} \sim 0.04$ Hz, which well reproduces the evolution of the dip in the $f_{\mathcal{S}}$ dstribution. $f_{\mathcal{S}}$ varies on timescales approximately 10 times shorter than $f_{\mathcal{V}}$, though both distributions vary over a similar range.}
\label{fig:f_cl_zoom}
\end{figure} 

To quantify and identify the origin of the non-axisymmetry we measure the clumping factor defined as 
\begin{equation}
f_{\rm{cl}} \equiv \frac{\langle \rho^2 \rangle}{\langle \rho \rangle ^2},
\label{eq:clump}
\end{equation}
where the angular brackets represent a spatial average. The average may be taken over different domains, and we outline our choices for the domain below. 

We first consider azimuthal clumping, 
\begin{equation}
f_{\phi}(r,\theta) = \frac{\langle \rho^2 \rangle_{\phi}}{\langle \rho \rangle ^2_{\phi}},
\end{equation}
where the spatial average is taken over $\phi$ but at fixed $(r,\theta)$ e.g., $\langle \rho \rangle_{\phi} = \int d \phi \ \rho$. Azimuthal clumping is a useful metric because $f_{\phi} \equiv 1$ for axisymmetric simulations and is therefore a local measurement of deviations from axisymmetry. In Fig. \ref{fig:azimuthal_clump} we plot the azimuthal clumping for the disc, base and wind point. In the disc and the wind, we find $f_{\phi} \approx 1$, indicating little deviation from axisymmetry. However, in the base $1 < f_{\phi} < 4$ consistent with our finding that over/under dense features are primarily at the base of the wind.     

Alternatively, we may consider clumping over a surface
\begin{equation}
f_{\mathcal{S}}(\phi) = \frac{\langle \rho^2 \rangle_{\mathcal{S}}}{\langle \rho \rangle ^2_{\mathcal{S}}},
\end{equation}
where the spatial average is over $r$ and $\theta$ but at fixed $\phi$ i.e. $\langle \rho \rangle_{\mathcal{S}} = \iint r^2 \sin \theta dr d\theta \ \rho$. We take the surface clumping over the region of the fast stream  $\mathcal{S} = \left\{ r_* < r < 10 \ r_*, 45^{\circ} < \theta < 65^{\circ}\right\}$. By this metric, even 2D axisymmetric winds are clumpy, see for example Fig. 2 of PSD98. 

We may also consider clumping over a volume
\begin{equation}
f_{\mathcal{V}} = \frac{\langle \rho^2 \rangle_{\mathcal{V}}}{\langle \rho \rangle ^2_{\mathcal{V}}},
\end{equation}
where the spatial average is over $r$, $\theta$ and $\phi$ i.e. $\langle \rho \rangle_{\mathcal{V}} = \iiint r^2 \sin \theta dr d\theta d\phi \ \rho$. We take the volume clumping over the region of the fast stream  $\mathcal{V} = \left\{ r_* < r < 10 \ r_*, 45^{\circ} < \theta < 65^{\circ}, 0 < \phi < \pi/4 \right\}$. For axisymmetric simulations $f_{\mathcal{S}} = f_{\mathcal{V}}$. 

In Fig. \ref{fig:f_cl_zoom}, we plot volume averaged clumping $f_{\mathcal{V}}$ (top panel) for the representative time $490 \ \rm{s} < t < 540 \ \rm{s}$ of the unperturbed (red, solid line) and perturbed (black, solid line) simulations. In the panels below, we plot the surface averaged clumping $f_{\mathcal{S}}$ at fixed $\phi = 0, \pi/16, \pi/8$ and $3\pi/16$. 

We find that the unperturbed wind has the highest volume averaged clumpiness over the duration of the outflow $200 \ s < t < 1000 \ s$, with $2 < f_{\mathcal{V}} < 9$, and it varies on timescales $t \sim 50 \ \rm{s}$. This same quantity for the perturbed run has a smaller range $2 < f_{\mathcal{V}} < 6$. The surface averaged clumpiness $f_{\mathcal{S}}$ exhibits the same range as the volume avergaed clumpiness, but varies on much shorter time scales $t \sim 2 \rm{s}$. 

The features in the $f_{\mathcal{S}}$ distribution vary with $\phi$ as a function of time. We plot three dashed diagonal lines, indicating the motion of three troughs in the clumpiness distribution. The diagonal lines indicate how a uniform sized feature moving at constant angular velocity propagates in time. In this case, we see a dip in $f_{\mathcal{S}}$ propagating for three orbits before dissipating. This corresponds to an approximate angular velocity $\Omega_{cl} \sim 0.04$ Hz. This is what one might expect for features of size $\Delta r = r_* \Delta \phi$ propagating at $v_K$. This is not to say that our result is resolution dependent, rather that the timescale of variations in $f_{\mathcal{S}}$ is set by the orbital velocity of clumps and not their outflow velocity. In the wind, the region where $\sigma(\rho)$ is maximum does in fact correspond to the part of the flow with this angular velocity ($r \lesssim 4 r_*$). The clumpiness of the wind is therefore dominated by material near the base of the wind,where the poloidal velocity is low. By contrast, $f_{\mathcal{V}}$ varies on timescales for material to be ejected from the wind $\tau_{\rm{wind}} \sim 10 r_*/ v_{\rm{wind}}$ where $v_{\rm{wind}} \sim 900 \rm{km/s}$. Disc perturbations can therefore cause variations of roughly a factor of 2 in $f_{\mathcal{S}}$, but on timescales about 10 times shorter than similar sized variations in $f_{\mathcal{V}}$ which are due to radial variations in the flow.

\subsubsection{Fourier Transform}
\begin{figure*}
                \centering
                \includegraphics[width=1.\textwidth]{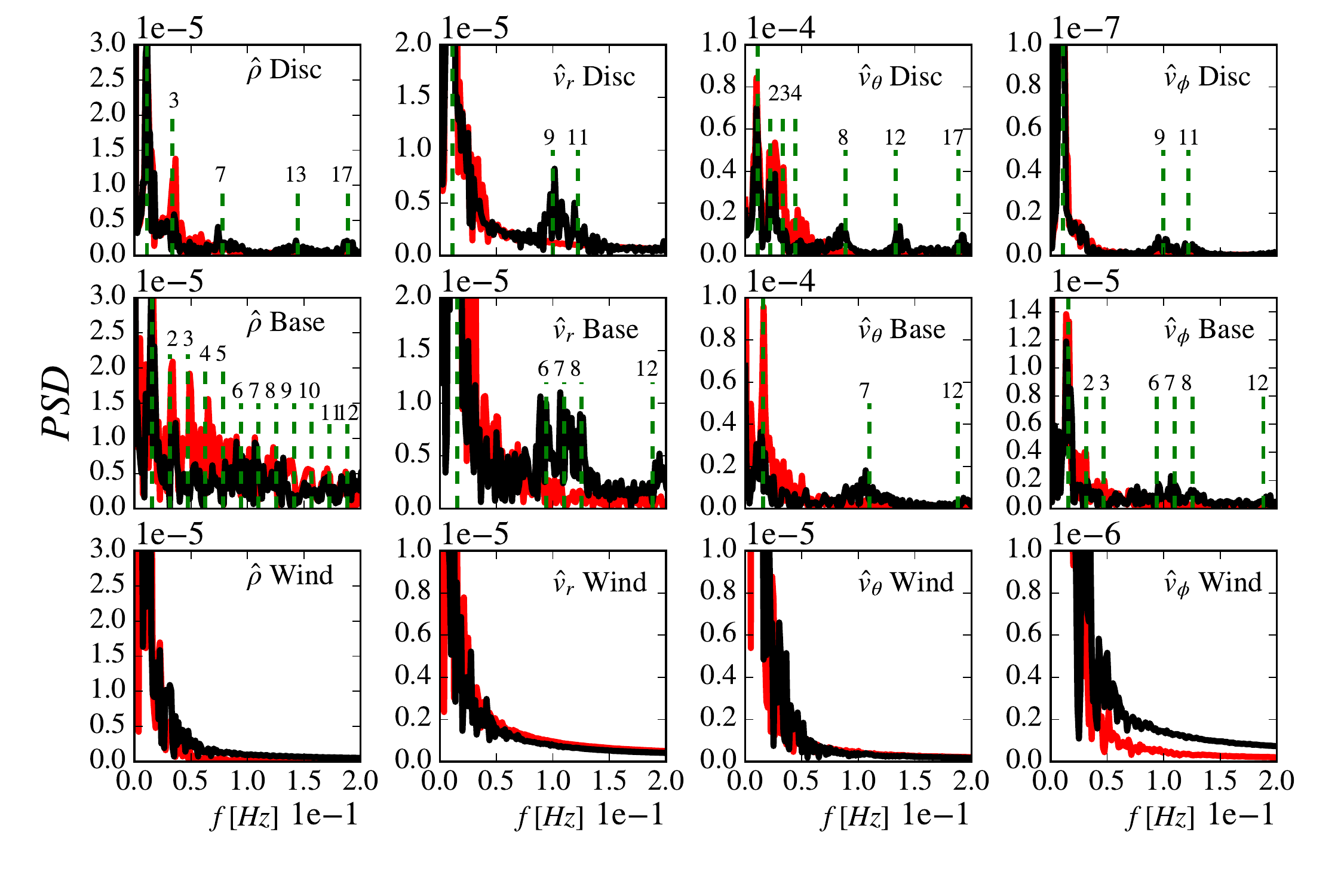}
        \caption{Normalized PSD for density $\hat{\rho}$ and each of the velocity components $\hat{v}_r$, $\hat{v}_{\theta}$ and $\hat{v}_{\phi}$ as a function of frequency at representative points in the disc, base and wind for the unperturbed (red) and perturbed (black) simulations. We indicate the $f_{1}$ mode (unlabeled green dashed line) for each point in the flow and mark the location of some easy to identify higher frequency modes. The wind does not have any easy to identify dominant modes.}
\label{fig:fourrier}
\end{figure*} 

To further investigate the time variability of the non-axisymmetry, we consider the Fourier transform, 
\begin{equation}
\hat{\rho}(\omega) = \int_{-\infty}^{\infty} \rho(t) e^{-2 \pi i \omega t} \ dt,
\end{equation}
where we apply a standard fast fourier transform (FFT) algorithm on our data. To eliminate any effects due to transient modes associated with the initial wind launching, we consider times $200 \leq t \leq 1000$ s, as the time during which an outflow is established. In Figure \ref{fig:fourrier}, we plot the normalized power spectrum distribution (PSD) for density $\hat{\rho}$ and each of the velocity components $\hat{v}_r$, $\hat{v}_{\theta}$ and $\hat{v}_{\phi}$ as a function of frequency $f$ for the disc, base and wind points. The normalization is chosen so that integrating the PSD over all frequencies equals unity.  

The PSD of the disc have the most well defined features, with the PSD of each quantity having a strong peak at fundamental frequency $f_1 = 0.8 f_K$, where $f_K$ is the local Keplerian frequency. We label higher multiples of this fundamental frequency via $f_n = n f_1$. The $\hat{\rho}$ PSD is dominated by a zero mode. The unperturbed simulation has prominent peaks at $f_1$ and $f_3$, while the perturbed solution has $f_1$,$f_3$,$f_7$, $f_{13}$ and $f_{17}$.  The $\hat{v}_r$ PSD has no zero mode and the unperturbed solution has peaks at $f_{1}$ and the perturbed solution at $f_{1}$, $f_{9}$ $f_{11}$. The $\hat{v}_{\theta}$ unperturbed PSD has peaks at $f_{1}$, $f_{2}$, $f_{3}$ and $f_{4}$ whereas the perturbed run at $f_{1}$,$f_{2}$,$f_{4}$,$f_{8}$, $f_{12}$ and $f_{17}$. The $\hat{v}_{\phi}$ has a dominant zero mode with the perturbed run having peaks at $f_{9}$ and $f_{11}$. The PSD being less noisy in the disc is consistent with the disc being in near hydrostatic equilibrium and maintaining the highest degree of axisymmetry.

The PSD for the base has the greatest number of modes with fundamental frequency $f_1 = 1.12 f_K$ .The unperturbed and perturbed runs have $\hat{\rho}$ PSD peaks at all modes $f_{1} \leq f_{n} \leq f_{12}$. The unperturbed $\hat{v}_r$ PSD has a $f_{1}$ mode and the perturbed run has $f_{1}$, $f_{6}$, $f_{7}$, $f_{8}$ and $f_{12}$. The unperturbed $\hat{v}_{\theta}$ PSD has a $f_{1}$ mode and the perturebd run $f_{1}$, $f_{7}$ and $f_{12}$. The unperturbed $\hat{v}_{\phi}$ PSD has a $f_{1}$, $f_{2}$ and $f_{3}$ mode and the perturbed run has $f_{1}$,$f_{6}$, $f_{7}$, $f_{8}$ and $f_{12}$ modes. The presence of very many modes is consistent with our finding that this is the most disordered part of the flow.

The wind PSD, all feature a large peak at low frequency, but unlike in the base and the disc it is hard to identify particular modes. The wind does not have any dominant modes because they are advected out of the domain.              

Certain modes may be present in our results because we have only simulated a wedge with $\Delta \phi = \pi/4$. Therefore we ran a test simulation with the full disc and concluded that this was not the case. A general feature is the perturbed wind has more modes, and these modes are at higher frequencies, than the unperturbed wind. A complete understanding of the modes present, even in the comparitively simple nearly hydrostatic disc is challenging and is left to future work.

\section{Discussion}
\label{sec:discussion}

\subsection{Form of Perturbation}
Here we present the first study of non-axisymmetric features of line driven disc winds. We restricted ourselves to vertical, subsonic, sinusoidal velocity perturbations of the form (\ref{eq:perturbation}), though other forms of perturbation are possible. We varied a few of the perturbation parameters, to check that our results are reasonably robust. In particular, we increased the magnitude of the perturbation $|\delta v|$ from 0.1 $c_s$ to 0.3 $c_s$. We also doubled the wavenumber of the perturbation $k_{\phi}$ i.e. from 8 to 16.
 
The overall conclusion from these experiments is that varying these parameters does not have a strong effect on the properties of the solution. Increasing the amplitude of the perturbation did not introduce any new modes. Doubling the wavenumber excited a $f_5$ mode in $v_{\phi}$ in the base and but kept other modes unchanged. This suggests the excited modes are the natural frequencies of the system and not due to the initial perturbation. After the inital mass outflow at $t \sim 200$ s, increasing the magnitude or changing the wavenumber led to a $\sim ~ 10\%$ change in the wind fluxes relative to our fiducial perturbation. This was a smaller difference than between our perturbed and unperturbed runs shown in Fig. \ref{fig:fluxes}. 

Removing the linear ramp up of the line driving led to $\sim 100 \%$ differences at early times, as the initial outburst occured at roughly $t \sim 100$ s. However, after this initial discrepency the flow settles down to the same $\sim ~ 10\%$ difference as seen in our other experiments. 
  
This suggests that for this form of the perturbation, the amount by which the gross outflow properties (mass, momentum and energy fluxes) can be varied is a factor of $\sim 2$. This is comparable to the variation we see in these quantities because of the time dependent nature of line driven disc winds.   

We also performed a simulation over the entire toroidal range $0 \leq \phi < 2\pi$ to ensure that simulating only a $\Delta \phi = \pi/4$ wedge would not introduce spurious modes into the solution. We did not find evidence that the wedge run had additional modes. In the $Q \approx dv/dz$ case this test was possible since the runtime was under two weeks running on 32 processors. When we relax this approximation this will be computationally expensive, so this null result is encouraging in this context.

The precise form of the perturbation does not seem to have an effect on the Fourrier modes, suggesting these are the natural modes of the system. Non-axisymmetric density features grow at the base of the wind, unlike in a persistent thermal wind where they are advected away. We speculate this growth is due to gas falling back on the disc after failing to launch. This scenario will be investigated in future work.

\subsection{Observations}

In the previous section, we have primarily discussed how a disc perturbation breaks the axisymmetry of the wind solution and argued that from the point of view of simulations these are non-trivial i.e $\mathcal{O}(1)$ effects. However, these axisymmetries are also observationally relevant.

First we point out that axisymmetric winds can result in non-axisymmetric observables. Velocity dependent observables (Doppler shifted lines) depend on the velocity component projected onto the line of sight. Because the line of sight typically breaks the natural symmetry of the system (axial symmetry in the case of disc winds), this leads to non-axisymmetric observables. This was done for example in the context of BAL QSOs where axisymmetric disc wind solutions generated non-axisymmetric, time dependent line profiles (Proga, Rodriguez-Hidalgo \& Hamann 2012). Therefore, we might expect the non-axisymmetric wind solutions we have found to further increase the time variability of these profiles.

Consider the emission measure of a particular line, that scales like $\sim \rho^2$. The maximum relative dispersion in $\phi$, Fig. \ref{fig:deviation_2d3d}, shows that the density varies by $\sim 3$. This would therefore lead to an increase in a factor of $\sim 10$ in the emisson measure at these points. These are of course over-estimates, because these are the points at which the non-axisymmetric features are strongest. However, it does suggest that if we are to calculate accurate synthetic line profiles we should use 3D wind solutions.

The effect on absorption measures of lines will be less pronounced since these vary $\sim \rho$. Therefore, we would expect these non-axisymmetric features to cause factors of at most $\sim 3$. Observationally such variations are challenging to measure because variations of this magnitude are seen in axisymmetric simulations, as a result of the winds inherent time dependence.

A key to disentangling these effects may be the timescales over which they operate. As an example, we found the clumpiness of the wind to vary on a time scale of $\sim 50$s, whereas the clumpiness along specific sightlines varied on $\sim 2$ s due to orbiting clumps at small radii. These timescales are too short to resolve for CV systems. However, for QSOs where the Keplerian timescale is about $10^3$ times longer variations due to clumpiness may be observable. Given that line emission depends sensitively on the locations of resonance points, it may be possible to see variations in the emission measure on the time scales of these clumps moving at Keplerian velocity. A detailed calculation of line profiles should be able to determine this.

\section{Conclusion}
\label{sec:conclusion}
We performed 3D simulations of line driven disc winds using the \textsc{Athena++} code where the velocity gradient was calculated using the $Q \approx dv/dz$ approximation. In the first 3D simulation we used axisymmetric initial conditions and found driven winds using $Q \approx dv/dz$ maintain axisymmetry and produce outflows consistent with results from 2D axisymmetric simulations.    

We then introduced a vertical, subsonic, non-axisymmetric velocity perturbation in the disc midplane. We compared the resulting outflow to the unperturbed simulation and found, the global properties of the outflow are largely unchanged, though fluxes may vary by up to 100 \% at any given time. The disc perturbation produces azimuthal clumping factors up to a factor of $\sim 4$ in the base of the wind. Clumpiness in the overall flow does not increase but varies on timescales much shorter than for axisymmetric winds.

The disc is least affected by the perturbation where deviations from axisymmetry $\delta_{\rm{max}}(\rho) \sim 10^{-3}$. In the wind base the deviation is largest with $\delta_{\rm{max}}(\rho) \sim 10^{-1}$ but decreases to $\delta_{\rm{max}}(\rho) \sim 10^{-2}$ in the fastest moving parts of the wind because any perturbations in the wind are carried out of the simulation domain. Non-axisymmetric features are, volume wise, primarily \emph{under-densities}, up to a factor of 3 lower than the toroidal average. There exist also \emph{over-dense} regions where densities are a factor of 3 higher than the toroidal average. These over/under dense regions exist primarily at the base of the wind, above the hydrostatic disc but upstream from the fastest moving parts of the wind. 

The non-axisymmetries have potential observational effects. We find regions with $\sim 3$ difference in densities, which leads to a factor of $\sim 10$ difference in the emission from these regions. The clumpiness of the flow is also found to vary on time scales of the orbital motion, rather than the outflow velocity for axisymmetric flows. This too may be potentially observable with sufficient time resolution.

We have found that a small initial perturbation leads to non-axisymmetries in the flow. However, these asymmetries tend to saturate and not grow without bounds. This suggests that in order for non-axisymmetries to grow beyond $\sim \rm{a \ few}$ as we have found, requires additional physics beyond what is included in this model.

We calculated the velocitiy shear in the approximation $Q \approx dv/dz$. In a later paper we will explore the case where this approximation is relaxed. This will lead to an azimuthal component of the radiation force, potentially changing the growth of non-axisymmetries beyond what we presented here. 

Another approach is where we account for the shielding of the wind by matter at the base of the wind. We found that most of the non-axisymmetric clumpiness is due to matter at small radii which is launched above the disc but fails to be expelled. This matter can shield the wind from ionizing radiation. This provides a mechanism to couple the base of the wind to the broader outflow, which our optically thin treatment does not capture. This requires a more proper treatment of the radiation transfer through the wind. We also plan on including radiation from the central object.

Finally we have assumed that the radiation field is axisymmetric. This may not be the case, as spiral structures can form due to accretion from a donor star in a CV system. Alternatively, perturbations driven by the magnetorotational instabiity, can generate non-axisymmetries in the radiation field.   

We conclude it is worthwhile to devote future efforts to simulating line driven disc winds in full 3D,  as non-axisymmetric effects may affect the flow and its observational features.    

\section*{Acknowledgements}
This work was supported by NASA under ATP grant NNX14AK44G. The authors also acknowledge useful discussions with Zhaohuan Zhu and Tim Waters.

\appendix

\section{Radiation Force Module}
\label{sec:appendix}
We describe our module for computing the line driving force for a disc and stellar radiation field.

\subsection{Coordinate System}

\begin{figure}
                \centering
                \includegraphics[width=0.45\textwidth]{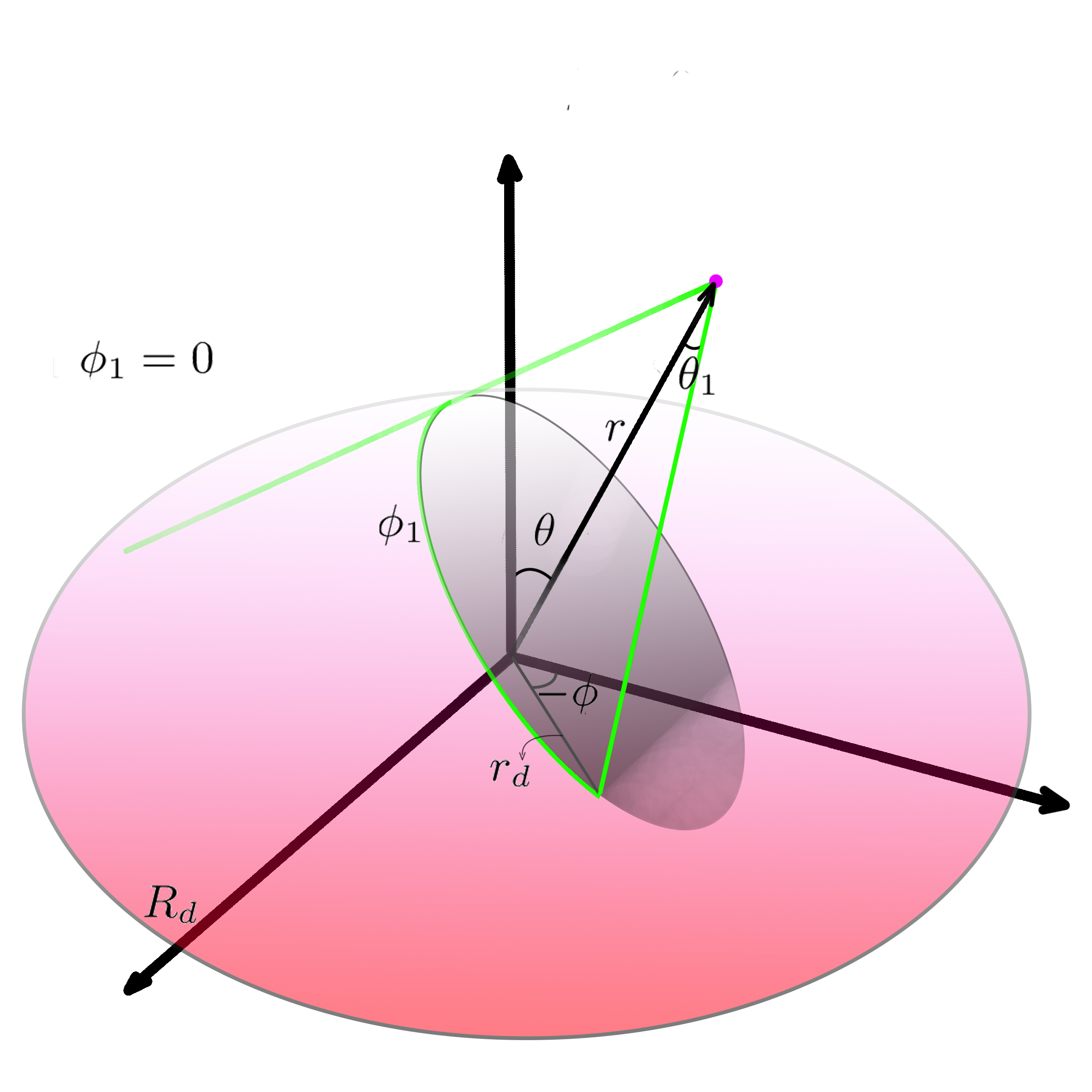}
        \caption{Summary of the coordinates used in our problem setup}
\label{fig:coordinates}
\end{figure} 
We use spherical coordinates $(r,\theta,\phi)$ to denote coordinates in the wind. These are the natural coordinates to use since these are the coordinates in the \textsc{Athena++} simulation. The disc is centered on the origin and lies in the XY plane with coordinates $(r_d,\pi/2,\phi)$ and inner and outer radii  $r_*$ and $R_d$ respectively. The star is also centered on the origin with radius $r_*$.  

Computing the radiation field requires integrating over the disc and star. PSD98 have shown that using cylindrical coordinates causes a loss of accuracy near the disc surface because of geometric foreshortening effects. PSD99 subsequently found that a spherical coordinate system $(r,\theta_1,\phi_1)$ centered on the wind point does not suffer from this issue so we use these coordinates as well. We choose the angular variable such that $\phi_1 = 0$ corresponds to points \emph{behind} the star. We show the different coodinates in Fig. \ref{fig:coordinates}. 

To compute the radiation force due to electron scattering (eq. \ref{eq:f_e}) and line driving (eq. \ref{eq:f_line}) we require the normal vector $\mathbf{n} = (\cos \theta_1, \sin \theta_1 \cos \phi_1, \sin \theta_1 \sin \phi_1 )$ and the solid angle $d \Omega = \sin \theta_1 d\theta_1 d\phi_1$. The intensity profile (eq. \ref{eq:intensity}) requires the disc radius

\begin{equation}
r_d = \sqrt{\bar{R}^2 + r^2 - 2 \bar{R}r \cos \theta_1},
\end{equation} 
where 
\begin{equation}
\bar{R} = \frac{r \cos \theta}{ \cos \theta_1 \cos \theta - \sin \theta_1 \cos \phi_1 \sin \theta}, 
\end{equation}
is the distance from the wind point to the disc point. It is useful to further perform the change of variables $\mu = \cos \theta_1$. 

\subsection{Numerical Integrator}

Our numerical scheme is inspired by the 3D integrator described in Press et al. (1988) but modified to integrate 2D functions. Suppose we are performing the integral
\begin{equation}
\iint f(x,y) \ dx dy,
\end{equation}
over a region $y_1(x) \leq y \leq y_2(x)$, $a \leq x \leq b$. We approximate this integral as 
\begin{equation}
\iint f(x,y) \ dx dy \approx \sum_{i=1}^{N} \sum_{j=1}^{M} f(x_i,y_j) w_j w_i,
\label{eq:numerical_integral}
\end{equation} 
where $w_i,w_j$ are weights, $x_i,y_j$ the appropriate abscissas and we are using $N,M$ integration points in the x and y directions respectively. The problem consists of choosing the weights and abscissas appropiately. In general we have found Gaussian quadrature performs best in terms of rapid convergence.

We specify the function $f(x,y)$, the y boundaries $y_1(x)$ and $y_2(x)$ and the x boundaries $a$ and $b$. In the first step, the algorithm chooses abscissas $x_i$ and weights $w_i$ in the interval $a \leq x_i \leq b$. Then for each $x_i$ the abscissas $y_j$ and weights $w_j$ are chosen in the interval $y_1(x_i) \leq y_j \leq y_2(x_i)$. The integral is then computed using (\ref{eq:numerical_integral}) where we integrate over the y variable first followed by the integral over x. For both the stellar and disc integrals we take $x = \mu$ and $y = \phi_1$ as our integration variables. We use $N_{\mu} \times N_{\phi_1} = 12 \times 12 = 144$ integration points. We artificillay break up the $\phi_1$ integral along $\phi_1 = \pi$ so as to avoid numerical asymetries in the $\phi$ direction.

For electron scattering (eq. \ref{eq:f_e}) and line driving using the $dv/dz$ approximation (eq. \ref{eq:f_line_dvdz}), we store the respective integration sum during the initialization step for each point in the wind. When computing the line driving force without approximation (eq. \ref{eq:f_line}), we store the components of $\mathbf{n}$ for each integration point and their corresponding weight. We need the directional vector to compute the Sobolev parameter and storing these allows us to not require calculating weights every timestep.  

\subsection{Testing}

We performed various tests to ensure our implementation of the radiation force was accurate to 1\%. We compared our numerical integrator to the analytic expressions for electron scattering for a uniform intensity $I = \rm{const.}$ and $I \sim r_d^{-2}$ ``Newtonian" disc from Feldmeier \& Shlosman (1999). We checked our implementation in the hydro by comparing to the 1D spherical CAK solution, as well as the CAK solution with finite disc correction (Pauldrach, Puls \& Kudritzki 1986; Gayley 1995). Finally we compared with results from PSD98, which explored cases with both stellar and disc radiation fields with Eddington parameter covering the range $0 < \Gamma_{d,s} < 1.18 \times 10^{-2}$.

Our tests led to us implementing two numerical features to ensure that the line driving is well behaved in \textsc{Athena++}, which were not necessary in the work of PSD98 and PSD99 who used the hydrodynamics code \textsc{Zeus 2D} (Stone \& Norman 1992). We imposed a ceiling on the strength of the line driving force. Whenever the momentum is to be updated due to the contribution from line driving, we ensure that the new velocity $v < 10 v_K(r_*)$ with $v_K$ the Keplerian velocity at radius $r_*$. We found that this prevents the formation of spuriously large velocities along the axis and is only triggered for the intial transient. However, we find late time solutions that have velocities $v < 10 \ v_K(r_*)$.

To further reduce transients, we turn on the radiation field using a linear ramp over $0 < t < 100$ s, which corresponds to roughly 5 inner disc orbits. These numerical features that have been implemented to ensure the line driving is well behaved but they are only necessary for high disc and stellar luminosities. i.e $\Gamma_d$ and $\Gamma_s$ approximately $10^{-2}$, which we do not consider in this work. However, we describe them here for completeness as we will consider higher luminosity cases in future studies.

\end{document}